\documentclass[aps,floatfix,showpacs,twocolumn,superscriptaddress]{revtex4} 
 
\usepackage{graphicx} 
\usepackage{dcolumn} 
\usepackage{amsmath}

\newcommand{\sfrac}[2]{\mbox{\footnotesize $\displaystyle \frac{#1}{#2}$}}

\begin{document} 
 
 
\title{%
$\;$\,\\[-5ex]
\hspace*{\fill}{\tt\normalsize ANL-PHY-10915-TH-2004, MPG-VT-UR~250/04}\\[1ex] %
Pseudoscalar Meson Radial Excitations} 
 
\author{A.\ H\"oll} 
\affiliation{Physics Division, Argonne National Laboratory, 
             Argonne, IL 60439-4843 U.S.A.} 
             
\author{A.\ Krassnigg}
\affiliation{Physics Division, Argonne National Laboratory, 
             Argonne, IL 60439-4843 U.S.A.} 
                          
\author{C.D.\ Roberts} 
\affiliation{Physics Division, Argonne National Laboratory, 
             Argonne, IL 60439-4843 U.S.A.} 
\affiliation{Fachbereich Physik, Universit\"at Rostock, D-18051 Rostock, 
Germany} 
             
\begin{abstract} 
\rule{0ex}{3ex} 
Goldstone modes are the only pseudoscalar mesons to possess a nonzero leptonic decay constant in the chiral limit when chiral symmetry is dynamically broken. The decay constants of their radial excitations vanish.  These features and aspects of their impact on the meson spectrum are illustrated using a manifestly covariant and symmetry preserving model of the kernels in the gap and Bethe-Salpeter equations.  
\end{abstract} 
 
\maketitle 
 
The meson spectrum contains three pseudoscalars [$I^G (J^P) L = 1^- (0^-) S$] with masses below $2\,$GeV \cite{pdg}: $\pi(140)$; $\pi(1300)$; and $\pi(1800)$.  Of these, the pion [$\pi(140)$] is well known and much studied.  The other two are observed, e.g., as resonances in the coherent production of three pion final states via pion-nucleus collisions \cite{experiment}.  In the context of a model constituent-quark Hamiltonian, these mesons are often viewed as the first three members of a $Q\bar Q$ $n\, ^1\!S_0$ trajectory, where $n$ is the principal quantum number; i.e., the $\pi(140)$ is the $S$-wave ground state and the others are its first two radial excitations.  By this reasoning the properties of the $\pi(1300)$ and $\pi(1800)$ are likely to be sensitive to details of the long range part of the quark-quark interaction because the constituent-quark wave functions will possess material support at large interquark separation.  Hence the development of an understanding of their properties may provide information about light-quark confinement, which complements that obtained via angular momentum excitations  \cite{a1b1}.

This view might reasonably be held about all trajectories of 
radial excitations.  However, the pseudoscalar trajectory is particularly interesting because its lowest mass member is QCD's Goldstone mode.  An explanation should therefore simultaneously describe: (1) chiral symmetry and its dynamical breaking; and (2) the possible correlation of the trajectory's higher mass members via an approximately linear radial Regge trajectory \cite{anisovich,norbury}.  Outcome (2) does not require that confinement in light-quark systems be expressed through the formation of a flux tube \cite{efimov}.  It is easily obtained in Poincar\'e invariant quantum mechanics \cite{klink} but requirement (1) is not.

A Poincar\'e covariant and symmetry preserving treatment of quark-antiquark bound states can be based on the Bethe-Salpeter equation (BSE)
\begin{equation}
\label{bse1}
\Gamma_{tu}(k;P) = \int^\Lambda_q [\chi(q;P)]_{sr}\, K_{rs}^{tu}(q,k;P)\,,
\end{equation}
where: $k$ is the relative and $P$ the total momentum of the constituents; $r$,\ldots,\,$u$ represent colour, Dirac and flavour indices; $\chi(q;P):= S(q_+) \Gamma(q;P) S(q_-)$, $q_\pm = q\pm P/2$; and $\int^\Lambda_q$ represents a translationally invariant regularisation of the integral, with $\Lambda$ the regularisation mass-scale \cite{mrt98,mr97}.  In Eq.\,(\ref{bse1}), $S$ is the renormalised dressed-quark propagator and $K$ is the fully amputated dressed-quark-antiquark scattering kernel.  The product $SS K$ is a renormalisation point invariant.  Hence, when the kernel is expressed completely in terms of renormalised Schwinger functions, the BSE's solution is independent of the regularisation mass-scale, which may be removed; viz., $\Lambda \to \infty$.

In a given channel the homogeneous BSE only has solutions for particular, separated values of $P^2$: $P^2=-m_n^2$, where $m_n$ is a bound state's mass,  whereat $\Gamma_n(k;P)$ is that bound state's Bethe-Salpeter amplitude.  (We use a Euclidean metric, with:  $\{\gamma_\mu,\gamma_\nu\} = 2\delta_{\mu\nu}$; $\gamma_\mu^\dagger = \gamma_\mu$; and $a \cdot b = \sum_{i=1}^4 a_i b_i$.  For a timelike vector $P_\mu$, $P^2<0$.)  In the flavour nonsinglet pseudoscalar channel the lowest mass solution is associated with the $\pi(140)$.  The homogeneous BSE next possesses a solution when $P^2$ assumes the value associated with the mass of the $\pi(1300)$.  This pattern continues so that in principle one may obtain the mass and amplitude of every pseudoscalar meson from Eq.\,(\ref{bse1}).  Herein we will illustrate this in practice for the two lowest-mass flavour-nonsinglet pseudoscalar mesons.

The dressed-quark propagator appearing in the BSE's kernel is determined by the renormalised gap equation 
\begin{eqnarray}
S(p)^{-1} & =&  Z_2 \,(i\gamma\cdot p + m^{\rm bm}) + \Sigma(p)\,, \label{gendse} \\
\Sigma(p) & = & Z_1 \int^\Lambda_q\! g^2 D_{\mu\nu}(p-q) \frac{\lambda^a}{2}\gamma_\mu S(q) \Gamma^a_\nu(q,p) , \label{gensigma}
\end{eqnarray}
where $D_{\mu\nu}$ is the dressed-gluon propagator, $\Gamma_\nu(q,p)$ is the dressed-quark-gluon vertex, and $m^{\rm bm}$ is the $\Lambda$-dependent current-quark bare mass.  The quark-gluon-vertex and quark wave function renormalisation constants, $Z_{1,2}(\zeta^2,\Lambda^2)$, depend on the renormalisation point, $\zeta$, the regularisation mass-scale and the gauge parameter.  The gap equation's solution has the form 
\begin{eqnarray} 
 S(p)^{-1} & = & i \gamma\cdot p \, A(p^2,\zeta^2) + B(p^2,\zeta^2) \,.
%
\label{sinvp} 
\end{eqnarray} 
It is obtained from Eq.\,(\ref{gendse}) augmented by the renormalisation condition
\begin{equation}
\label{renormS} \left.S(p)^{-1}\right|_{p^2=\zeta^2} = i\gamma\cdot p +
m(\zeta)\,,
\end{equation}
where $m(\zeta)$ is the renormalised mass: 
\begin{equation}
Z_2(\zeta^2,\Lambda^2) \, m^{\rm bm}(\Lambda) = Z_4(\zeta^2,\Lambda^2) \, m(\zeta)\,,
\end{equation}
with $Z_4$ the Lagrangian mass renormalisation constant.  In QCD the chiral limit is unambiguously defined by
\begin{equation}
\label{limchiral}
Z_2(\zeta^2,\Lambda^2) \, m^{\rm bm}(\Lambda) \equiv 0 \,, \forall \Lambda \gg \zeta \,,
\end{equation}
which is equivalent to stating that the renormalisation-point-invariant current-quark mass $\hat m = 0$.

The $1^- (0^-) S$ ($n\, ^1\!S_0$) trajectory contains the pion, whose properties are fundamentally governed by the phenomenon of dynamical chiral symmetry breaking (DCSB).  One expression of the chiral properties of QCD is the axial-vector Ward-Takahashi identity 
\begin{eqnarray}
\nonumber
P_\mu \Gamma_{5\mu}^j(k;P) & =& S^{-1}(k_+) i \gamma_5\frac{\tau^j}{2}
+  i \gamma_5\frac{\tau^j}{2} S^{-1}(k_-)\\
&& - \, 2i\,m(\zeta) \,\Gamma_5^j(k;P) ,
\label{avwtim}
\end{eqnarray}
which we have here written for two quark flavours, each with the same current-quark mass: $\{\tau^i:i=1,2,3\}$ are flavour Pauli matrices.  In Eq.\,(\ref{avwtim}), $\Gamma_{5\mu}^j(k;P)$ is the axial-vector vertex:  
\begin{eqnarray}
\nonumber
\left[\Gamma^j_{5\mu}(k;P)\right]_{tu}
 & = &  Z_2 \left[\gamma_5\gamma_\mu \frac{\tau^j}{2} \right]_{tu}\\
 &&  + \int^\Lambda_q
[\chi^j_{5\mu}(q;P)]_{sr} K_{tu}^{rs}(q,k;P)\,,
\label{avbse}
\end{eqnarray}
and $\Gamma_5^j(k;P)$ is the pseudoscalar vertex
\begin{eqnarray}
\nonumber
\left[\Gamma_{5}(k;P)\right]_{tu}
 & = &  Z_4 \left[\gamma_5 \frac{\tau^j}{2}\right]_{tu}\\
 &&  + \int^\Lambda_q
[\chi^j_{5}(q;P)]_{sr} K_{tu}^{rs}(q,k;P)\,.
\label{psbse}
\end{eqnarray}
 
The quark propagator, axial-vector and pseudoscalar vertices are all expressed via integral equations; i.e., Dyson-Schwinger equations (DSEs).  Equation~(\ref{avwtim}) is an exact statement about chiral symmetry and the pattern in which it is broken. Hence it must always be satisfied.  Since that cannot be achieved veraciously through fine tuning, the distinct kernels of Eqs.\,(\ref{gendse}), (\ref{gensigma}), (\ref{avbse}), (\ref{psbse}) must be intimately related.  Any theoretical tool employed in calculating properties of the pseudoscalar and pseudovector channels must preserve that relationship if the results are to be both quantitatively and qualitatively reliable.

A weak coupling expansion of the DSEs yields perturbation theory and satisfies this constraint.  However, that truncation scheme is not useful in the study of bound states nor other intrinsically nonperturbative phenomena.  Fortunately at least one nonperturbative systematic and symmetry preserving scheme exists. (Reference\,\cite{mandarvertex} gives details).  This entails that the full implications of Eq.\,(\ref{avwtim}) can be both elucidated and illustrated.

Every flavour nonsinglet pseudoscalar meson is exhibited as a pole contribution to the axial-vector and pseudoscalar vertices \cite{mrt98}: viz.,
\begin{eqnarray}
\left. \Gamma_{5 \mu}^j(k;P)\right|_{P^2+m_{\pi_n}^2 \approx 0}&=&   \frac{f_{\pi_n} \, P_\mu}{P^2 + 
m_{\pi_n}^2} \Gamma_{\pi_n}^j(k;P)\,, \label{genavv} \\
\left. i\Gamma_{5 }^j(k;P)\right|_{P^2+m_{\pi_n}^2 \approx 0}
&=&   \frac{\rho_{\pi_n} }{P^2 + 
m_{\pi_n}^2} \Gamma_{\pi_n}^j(k;P)\,, \label{genpv} 
\end{eqnarray}
wherein we have omitted terms regular in the neighbourhood of the pole, $\Gamma_{\pi_n}^j(k;P)$ is the bound state's canonically normalised Bethe-Salpeter amplitude: 
\begin{eqnarray} 
\nonumber
\lefteqn{
\Gamma_{\pi_n}^j(k;P) = \tau^j \gamma_5 \left[ i E_{\pi_n}(k;P) + \gamma\cdot P F_{\pi_n}(k;P) \right. }\\
&+& \left.
    \gamma\cdot k \,k \cdot P\, G_{\pi_n}(k;P) + 
\sigma_{\mu\nu}\,k_\mu P_\nu \,H_{\pi_n}(k;P)  \right] \! ; \label{genpibsa} \end{eqnarray}
and
\begin{eqnarray} 
\label{fpin} f_{\pi_n} \,\delta^{ij} \,  P_\mu &=& Z_2\,{\rm tr} \int^\Lambda_q 
\sfrac{1}{2} \tau^i \gamma_5\gamma_\mu\, \chi^j_{\pi_n}(q;P) \,, \\
\label{cpres} i  \rho_{\pi_n}\!(\zeta)\, \delta^{ij}  &=& Z_4\,{\rm tr} 
\int^\Lambda_q \sfrac{1}{2} \tau^i \gamma_5 \, \chi^j_{\pi_n}(q;P)\,.
\end{eqnarray} 
The residues expressed exactly in Eqs.\,(\ref{fpin}) and (\ref{cpres}), are gauge invariant and cutoff independent.  

For a structureless pseudoscalar meson, $F_{\pi_n}(k;P)\equiv 0 \equiv G_{\pi_n}(k;P) \equiv H_{\pi_n}(k;P)$ in Eq.\,(\ref{genpibsa}).  The first two of these functions can be described as characterising the pseudoscalar meson's pseudovector components; and the last, its pseudotensor component.  The associated Dirac structures necessarily occur in a Poincar\'e covariant bound state description and they signal the presence of quark orbital angular momentum.

Equation\,(\ref{avwtim}) combined with (\ref{genavv}) -- (\ref{cpres}) yields \cite{mrt98,mr97}
\begin{equation}
\label{gmorgen} f_{\pi_n} m_{\pi_n}^2 = 2 \, m(\zeta)  \, 
\rho_{\pi_n}(\zeta)\,; 
\end{equation}
i.e., an identity valid for every $0^-$ meson and irrespective of the magnitude of the current-quark mass \cite{mishasvy}.  

We focus now on the ground state $n=0$ pseudoscalar meson and assume that all pseudoscalar excitations are more massive.  In this case DCSB in QCD
entails \cite{mrt98}, via Eqs.\,(\ref{sinvp}),(\ref{avwtim}) and (\ref{genavv}) -- (\ref{cpres}), 
\begin{equation}
\label{gt1}
f_{\pi_0}^0 \, E_{\pi_0}(k;0) = B(k^2,\zeta^2)\,,
\end{equation}
where $f_{\pi_n}^0 := \lim_{\hat m \to 0}\, f_{\pi_n}$, from which follows \begin{equation}
\rho_{\pi_0}^0(\zeta) := \lim_{\hat m \to 0}\,\rho(\zeta) = -\frac{1}{f^0_{\pi_0} } \langle \bar q q \rangle^0_\zeta\,,
\end{equation}
wherein  
\begin{equation} 
\label{qbq0} \,-\,\langle \bar q q \rangle_\zeta^0 = \lim_{\Lambda\to \infty} 
Z_4(\zeta^2,\Lambda^2)\, N_c \, {\rm tr}_{\rm D}\int^\Lambda_q\!
S^{0}(q,\zeta)\,,  
\end{equation} 
is the vacuum quark condensate \cite{langfeld}.  The scalar functions in Eq.\,(\ref{sinvp}) that characterise the dressed-quark propagator are both positive definite at spacelike momenta \cite{latticequark,mandar}.  This fact along with their ultraviolet behaviour in the chiral limit \cite{masslessuv} guarantee $0 < -\langle \bar q q \rangle^0_\zeta< \infty$.  Furthermore, Eq.\,(\ref{gt1}) and related quark-level Goldberger-Treiman relations \cite{mrt98} involving $A(k^2;\zeta^2)$ ensure that  $0<f^0_{\pi_0}<\infty$.  Hence $m_{\pi_0}=0$ in the chiral limit, and in the neighbourhood of this limit the so-called Gell-Mann--Oakes--Renner relation is a corollary of Eq.\,(\ref{gmorgen}).

We next consider the $n>0$ pseudoscalar mesons: $m_{\pi_{n> 0}}>m_{\pi_0}$ by assumption, and hence $m_{\pi_{n> 0}} \neq 0$ in the chiral limit.  The existence of a bound state entails that $\chi_n(k;P)$ is a finite matrix-valued function.  Moreover, the ultraviolet behaviour of the quark-antiquark scattering kernel in QCD guarantees that Eq.\,(\ref{cpres}) is cutoff independent.  Thus
\begin{equation}
\rho_{\pi_{n}}^0(\zeta):= \lim_{\hat m\to 0} \rho_{\pi_{n}}(\zeta) <\infty \,, \; \forall \, n\,.
\end{equation}
Hence, it is a necessary consequence of chiral symmetry and its dynamical breaking in QCD; viz., Eq.\,(\ref{gmorgen}), that
\begin{equation}
\label{fpinzero}
f_{\pi_n}^0 \equiv 0\,, \forall \, n\geq 1\,.
\end{equation}
This result is consistent with Refs.\,\cite{dominguez}, as appreciated in Ref.\,\cite{volkov}.

This argument is legitimate in any theory that has a valid chiral limit.  It is logically possible that such a theory does not exhibit DCSB; i.e., realises chiral symmetry in the Wigner mode.  This is the case, e.g., in QCD above the critical temperature for chiral symmetry restoration \cite{bastirev}.  Equation (\ref{gmorgen}) is still valid in the Wigner phase.  However, its implications are different.  

Without DCSB; namely, in the Wigner phase, one has 
\begin{equation}
\label{Bm0}
B^W(0,\zeta^2)\propto m(\zeta) \propto\hat m\,;
\end{equation}
i.e., the constituent-quark mass vanishes in the chiral limit.  This result is accessible in perturbation theory.  Equation (\ref{gt1}) applies if there is a massless bound state in the chiral limit.  Suppose such a bound state persists in the absence of DCSB.  (If that is false then considering this particular case is unnecessary.  However, it is true at the transition temperature in QCD \cite{bastirev}.) It then follows from Eqs.\,(\ref{gt1}) and (\ref{Bm0}) that \begin{equation}
f^W_{\pi_0} \propto \hat m\,.
\end{equation}
In this case the leptonic decay constant of the ground state pseudoscalar meson also vanishes in the chiral limit.  

It is always true that 
\begin{equation}
f_{\pi_0}\,\rho_{\pi_{0}}(\zeta) \stackrel{\hat m\approx 0}{\propto} - \langle \bar q q \rangle_\zeta^0 \,.
\end{equation}
In the Wigner phase \cite{langfeld}, $ \langle \bar q q \rangle^{0\,W}_\zeta \propto \hat m^3$.  Hence, via Eq.\,(\ref{gmorgen}), if a rigorously chirally symmetric theory possesses a massless pseudoscalar bound state then
\begin{equation}
m^W_{\pi_0} \stackrel{\hat m\approx 0}{\propto} \hat m \,.
\end{equation}
In this case there is also a degenerate scalar meson partner whose mass behaves in the same manner.

We have elucidated exact results.  They can be illustrated accurately in a model that both preserves QCD's ultraviolet properties and exhibits DCSB.  For this purpose we employ the renormalisation-group-improved (RGI) rainbow-ladder DSE model introduced in Ref.\,\cite{maristandy1} whose widespread application is reviewed in Ref.\,\cite{revpieter}.  The heart of the model is an \textit{Ansatz} for the Bethe-Salpeter kernel in Eq.\,(\ref{bse1}):
\begin{eqnarray}
\nonumber \lefteqn{
K^{tu}_{rs}(q,k;P) = - \,{\cal G}((k-q)^2) }\\
&&  \times \, D_{\mu\nu}^{\rm free}(k-q)\,\left[\gamma_\mu \frac{\lambda^a}{2}\right]_{ts} \, \left[\gamma_\nu \frac{\lambda^a}{2}\right]_{ru} \!, \label{ladderK}
\end{eqnarray}
wherein $D_{\mu\nu}^{\rm free}(\ell)$ is the free gauge boson propagator and
\begin{eqnarray}
\nonumber \lefteqn{
\frac{{\cal G}(s)}{s} = \frac{\pi}{\omega^6} \, D\, s\, {\rm e}^{-s/\omega^2}}\\
&& + \frac{2\pi \gamma_m}{\ln\left[ \tau + \left(1+s/\Lambda_{\rm QCD}^2\right)^2\right]} \, {\cal F}(s)\,,
\label{Gkmodel}
\end{eqnarray}
with ${\cal F}(s)= [1-\exp(-s/[4 m_t^2])]/s$, $m_t=0.5\,$GeV, $\tau={\rm e}^2-1$, $\gamma_m=12/25$ and $\Lambda_{\rm QCD} = \Lambda^{(4)}_{\overline{MS}} = 0.234$.  

This form expresses the interaction as a sum of two terms.  The second ensures that perturbative behaviour is correctly realised at short range; namely, as written, for $(k-q)^2 \sim k^2 \sim q^2 \gtrsim 1 - 2\,$GeV$^2$, $K$ is precisely as prescribed by QCD.  On the other hand, the first term in ${\cal G}(k^2)$ is a model for the long-range behaviour of the interaction.  It is a finite width representation of the form introduced in Ref.\,\cite{mn83}, which has been rendered as an integrable regularisation of $1/k^4$ \cite{mm97}.  This interpretation, when combined with the result that in a heavy-quark--heavy-antiquark BSE the RGI ladder truncation is exact \cite{mandarvertex}, is consistent with ${\cal G}(k^2)$ leading to a Richardson-like potential \cite{richardson} between static sources.

The active parameters in Eq.\,(\ref{Gkmodel}) are $D$ and $\omega$, which together determine the integrated infrared strength of the rainbow-ladder kernel, but they are not independent.  In fitting a selection of observables \cite{maristandy1}, a change in one is compensated by altering the other; e.g., on the domain $\omega\in[0.3,0.5]\,$GeV, the fitted observables are approximately constant along the trajectory $\omega D = (0.72 \, {\rm GeV})^3 =: m_g^3$.  Herein we use $\omega= 0.38\,$GeV.

\begin{figure}[t] 
 
\centerline{\includegraphics[width=0.38\textwidth,angle=-90]{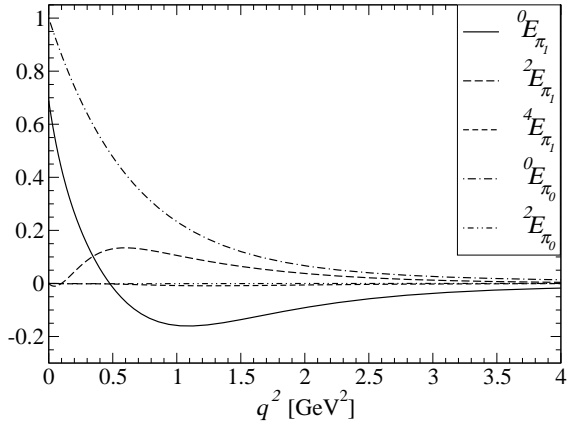}}
 
\caption{\label{Fig1} Dimensionless low-order Chebyshev moments of the scalar function that characterises the dominant amplitude in Eq.\,(\protect\ref{genpibsa}), for the ground ($\pi_0$) and first excited ($\pi_1$) states.}
\end{figure} 

With a given truncation of the BSE's kernel there is a unique kernel for the gap equation which ensures the axial-vector Ward-Takahashi identity, Eq.\,(\ref{avwtim}), is automatically satisfied \cite{mandarvertex}.  This partner to Eq.\,(\ref{ladderK}) is a rainbow gap equation; i.e., Eq.\,(\ref{gendse}) with 
\begin{equation} 
%
\Sigma(p)=\int^\Lambda_q\! {\cal G}((p-q)^2) D_{\mu\nu}^{\rm free}(p-q) \frac{\lambda^a}{2}\gamma_\mu S(q) \frac{\lambda^a}{2}\gamma_\nu . \label{rainbowdse} 
\end{equation} 

We have calculated properties of the ground and first excited states.  The first step was to solve the gap equation, Eq.\,(\ref{rainbowdse}), for a specified renormalised current-quark mass.  With the dressed-quark propagator thus obtained, solutions of the homogeneous BSE, Eq.\,(\ref{bse1}) with (\ref{ladderK}), were obtained via a straightforward numerical procedure that yields every scalar function in Eq.\,(\ref{genpibsa}).  The general procedure is described in detail in Ref.\,\cite{mr97}, and Ref.\,\cite{kr03} explains the method necessary to isolate excited states.

In Fig.\,\ref{Fig1} we depict the lowest Chebyshev moments of the pseudoscalar amplitude in Eq.\,(\ref{genpibsa}); i.e., $^{0,2} \! E_{\pi_{0}}(k^2)$ and $^{0,2,4} \! E_{\pi_{1}}(k^2)$, where 
\begin{eqnarray}
\nonumber
\lefteqn{^i \! E_{\pi_{0,1}}(k^2) = } \\
&&\frac{2}{\pi}\int_0^\pi\! d\beta \; \sin^2\!\beta \, U_i(\cos\beta)\, E_{\pi_{0,1}}(k^2,k\cdot P;P^2)\,. \label{cheb}
\end{eqnarray}
with $U_i(x)$ a Chebyshev polynomial of the second kind and $k\cdot P := \cos\beta \sqrt{k^2 P^2}$.  The odd moments, $i=1,3,5$,\ldots, etc., vanish in the case of equal mass constituents.  The Chebyshev moments are obtained from the canonically normalised Bethe-Salpeter amplitudes but, for illustrative simplicity, the functions depicted are rescaled by the positive constant $^{0} \! E_{\pi_{0}}(k^2=0)$.  

\begin{figure}[t] 
 
\centerline{\includegraphics[width=0.38\textwidth,angle=-90]{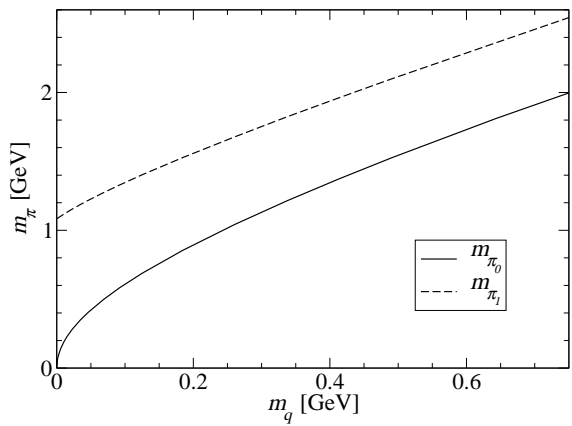}}
 
\caption{\label{Fig2} Evolution of the mesons' masses with renormalised current-quark mass, $m_q=m(\zeta=1\,{\rm GeV})$.}
\end{figure} 

All Chebyshev moments of $E_{\pi_1}$ possess a single zero, whereas those of $E_{\pi_0}$ exhibit none.  This similarity to the wave functions of radial excitations in quantum mechanics is not particular to manifestly covariant BSE studies \cite{metsch2,bakker}.  It is evident that the zeroth Chebyshev moment almost completely determines the pseudoscalar amplitude in the ground state pseudoscalar meson.  For the first excited state, however, the second moment is also required to obtain a good approximation to $E_{\pi_1}(k;P)$.  The pseudovector and pseudotensor amplitudes are nonzero in the ground and first excited states, and they are materially important in the calculation of their properties.  The bulk qualitative features of the scalar functions characterising these amplitudes are the same as those described in connection with the pseudoscalar amplitude. 

\begin{figure}[t] 
 
\centerline{\includegraphics[width=0.38\textwidth,angle=-90]{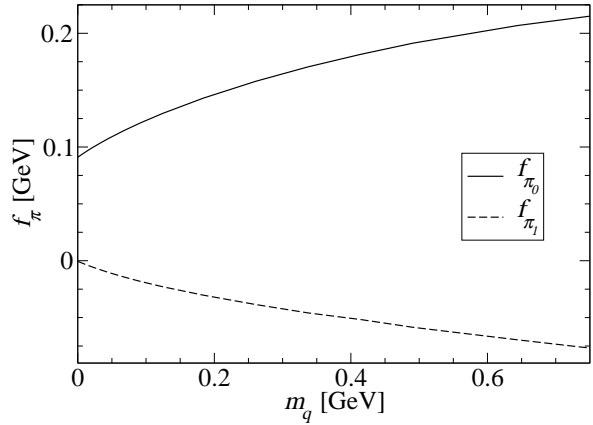}}
 
\caption{\label{Fig3} Evolution of the mesons' leptonic decay constants with renormalised current-quark mass, $m_q=m(\zeta=1\,{\rm GeV})$.}
\end{figure}

In Fig.\,\ref{Fig2} we depict the evolution with current-quark mass of the masses of the ground and first excited pseudoscalar states, and in Fig.\,\ref{Fig3} we display the behaviour of the leptonic decay constants.  Calculated results at points of particular interest on these trajectories are presented in Table \ref{tablea}.  NB.\ The ladder truncation supports ideal mixing.  This is not a good picture of ground state $s\bar s$ pseudoscalars, for which $s$-channel gluon (OZI suppressed) contributions are important.  (See, e.g., Ref.\,\cite{klabucar}.)  However, it should prove increasingly reliable as the current-quark mass increases, since the RGI ladder truncation is exact in the static source limit, and/or as the mass of the bound state increases because this mass-scale, too, suppresses diagrams that violate the OZI rule.

The chiral behaviour we predicted for the leptonic decay constant of the pseudoscalar meson excited states is exemplified in Fig.\,\ref{Fig3}: viz., $f_{\pi_1}=0$ for $\hat m = 0$.  It is notable that this decay constant is negative as the current-quark mass increases away from the chiral limit.  Hence the residue in the pseudovector vertex of the pole associated with the $2 ^1\!S_0$ meson is negative on $\hat m>0$.  It is evident from the table that the residue at the associated pole in the pseudoscalar vertex, $\rho_{\pi_1}(\zeta)$, is also negative.  Thus is Eq.\,(\ref{gmorgen}) satisfied.  We anticipate that the pole residues alternate in sign; i.e., they are positive for even $n$ but negative for odd $n$.  This is required of a \textit{vertex} that, considered as a function of $P^2$, is continuous and does not vanish between adjacent bound state poles.  Since it is the square of the residues that appear in the scalar functions which characterise the axial-vector and pseudoscalar vacuum polarisations: $\Pi_A(P^2)$, $\Pi_P(P^2)$, this feature is unlikely to have readily observable consequences.

\begin{table}[t] 
\caption{\label{tablea} Calculated results for properties of the ground and first excited state pseudoscalar mesons: $m(\zeta_0):= \hat m/(\ln\zeta_0/\Lambda_{\rm QCD})^{\gamma_m}$, $\zeta_0=1\,$GeV; $\zeta=19\,$GeV.  Available experimental values (in GeV) \cite{pdg}: $m_{\pi_0}=0.14\,$, $m_{\pi_1}=1.3\pm 0.1\,$; $m_\eta= 0.547$, $m_{\eta(2S)}=1.293\pm 0.005$; 
$f_{\pi_0}=0.092\,$.  Dimensions in table are GeV except for $\rho(\zeta)$, which is listed in GeV$^2$.\vspace*{1ex}}
\begin{ruledtabular} 
\begin{tabular*} 
{\hsize} {l@{\extracolsep{0ptplus1fil}} 
l@{\extracolsep{0ptplus1fil}}
l@{\extracolsep{0ptplus1fil}}l@{\extracolsep{0ptplus1fil}}
l@{\extracolsep{0ptplus1fil}}l@{\extracolsep{0ptplus1fil}}
l@{\extracolsep{0ptplus1fil}}} 
$m(\zeta_0)$ & $m_{\pi_0}$ & $m_{\pi_1}$ & $f_{\pi_0}$ & $f_{\pi_1}$ & $\rho_{\pi_0}(\zeta)$ & $\rho_{\pi_1}(\zeta)$ \\
$0$ & $0$ & $1.08$ & $0.091$ & $\;\;\;0$ & $(0.51)^2$ & $- (0.49)^2$ \\
$0.0055$ & $0.14$ & $1.10$ & $0.093$ & $-0.002$ & $(0.52)^2$& $- (0.49)^2$\\
$0.125$ & $0.69$ & $1.40$ &$0.130$ & $-0.023$& $(0.64)^2$& $- (0.54)^2$
%
\end{tabular*} 
\end{ruledtabular} 
\end{table} 

The parameter $\omega$ in Eq.\,(\ref{Gkmodel}) defines a length-scale $r_a = 1/\omega$, and magnifying $r_a$ increases the range of strong attraction in the model.  In our calculations we found, as anticipated, that the properties of the pion's first radial excitation are sensitive to the long-range part of the interaction.  For example, while a $30$\% increase in $r_a$ raises $m_{\pi_0}$ by merely $3$\%, it reduces $m_{\pi_1}$ by $14$\%.  In contrast, on the domain of current-quark masses for which our present calculations are reliable, the results are consistent with $m_{\pi_0}/m_{\pi_1}\to 1^-$ in the heavy-quark limit.  

We estimated the charge radius of the pion and its first radial excitation using the symmetry preserving impulse approximation introduced in Ref.\,\cite{cdrpion} with the Ball-Chiu \textit{Ansatz} \cite{bc80} for the dressed-quark-photon vertex: $r_{\pi_1} = 1.7\,r_{\pi_0}$.  While a more reliable calculation would employ a self-consistently calculated dressed-quark-photon vertex \cite{pmpion}, our estimate for the ratio should be a fair guide.

It is natural to extend our quantitative analysis to: larger current-quark masses; higher radial excitations; and systems in which the constituents have different masses.  This expansion of the domain on which the manifestly covariant and symmetry preserving model is applied will inform our interpretation of observables, and guide the model's improvement and further employment.  It will not, however, alter the exact results the model has been used to illustrate.  They serve as a constructive constraint on approaches advocated for the study of (exotic) hadron spectroscopy and interactions.   

\begin{acknowledgments} 
We acknowledge interactions with P.\ Maris, P.\,C.\ Tandy, M.\,K.\ Volkov and V.\,L.\ Yudichev.
This work was supported by: Austrian Research Foundation \textit{FWF, 
Erwin-Schr\"odinger-Stipendium} no.\ J2233-N08; Department of Energy, 
Office of Nuclear Physics, contract no.\ W-31-109-ENG-38; National Science Foundation contract no.\ INT-0129236; the \textit{A.\,v.\ 
Humboldt-Stiftung} via a \textit{F.\,W.\ Bessel Forschungspreis};  and benefited from the ANL Computing Resource Center's facilities.
\vspace*{\fill}
\end{acknowledgments} 

\end{document}